\documentclass[preprint,floatfix,showpacs]{revtex4}%
\usepackage{graphicx}
\usepackage{amsmath,amssymb}
\usepackage{amsmath}
\usepackage{amsfonts}
\usepackage{amssymb}%
\def\c5q{c^5_q}

\def\pl2{\ell^2}
\def\pq2{q^2}

\begin{document}
\title{Form factors of radiative pion decays in nonlocal chiral quark models}
\author{D. G\'{o}mez Dumm$^{a,b}$, S. Noguera$^{c}$ and N. N. Scoccola$^{b,d,e}$}
\affiliation{$^{a}$ IFLP, CONICET $-$ Dpto.\ de F\'{\i}sica, Universidad
Nacional de La Plata, C.C. 67, (1900) La Plata, Argentina.}
\affiliation{$^{b}$ CONICET, Rivadavia 1917, (1033) Buenos Aires,
Argentina.} \affiliation{$^{c}$ Departamento de F\'{\i}sica Te\'orica and
Instituto de F\'{\i}sica Corpuscular, Universidad de Valencia-CSIC,
E-46100 Burjassot (Valencia), Spain.} \affiliation{$^{d}$ Physics
Department, Comisi\'on Nacional de Energ\'{\i}a At\'omica, }
\affiliation{Av.\ Libertador 8250, (1429) Buenos Aires, Argentina}
\affiliation{$^{e}$ Universidad Favaloro, Sol{\'\i}s 453, (1078) Buenos
Aires, Argentina.}

\begin{abstract}
We study the radiative pion decay $\pi^{+}\rightarrow e^{+}\nu_{e}\gamma$
within nonlocal chiral quark models that include wave function
renormalization. In this framework we analyze the momentum dependence of the
vector form factor $F_{V}(q^{2})$, and the slope of the axial-vector form
factor $F_{A}(q^{2})$ at threshold. Our results are compared with available
experimental information and with the predictions given by the NJL model.
In addition we calculate the low energy constants $\ell_5$ and $\ell_6$,
comparing our results with the values obtained in chiral perturbation
theory.
\end{abstract}

\maketitle

\section{Introduction}

The radiative pion decay $\pi^{+}\rightarrow e^{+}\nu_{e}\gamma$
is a very interesting process from different points of view.
According to the standard description, the corresponding decay
amplitude consists of the inner bremsstrahlung (IB) and
structure-dependent (SD) terms. The former can be associated with
the diagrams in which the photon is radiated by the electrically
charged external legs (either pion or lepton), while the SD terms
correspond to the photon emission from intermediate states
generated by strong interactions. Since the IB contribution turns
out to be helicity suppressed, this process is an adequate channel
to study the SD amplitude, which provides information about strong
interactions in the nonperturbative regime.

The SD contribution can be parameterized through the introduction of
vector and axial-vector form factors, $F_{V}(q^{2})$ and $F_{A}(q^{2})$
respectively, where $q^{2}$ is the squared invariant mass of the $e^{+}\nu
_{e}$ pair~\cite{Bryman:1982et}. From the experimental point of view,
recent analyses~\cite{Bychkov:2008ws} of $\pi^{+} \rightarrow e^{+}
\nu_{e} \gamma$ decays have allowed the measurement of $F_{V}(0)$,
$F_{A}(0)$, and the slope of the form factor $F_V(q^2)$ at $q^2=0$. In
addition, ongoing experiments are expected to reach enough statistics to
determine the slope of $F_A(q^2)$ at $q^2=0$ in the near future. On the
theoretical side, the analysis of the form factors has been carried out in
the framework of Chiral Perturbation Theory ($\chi$PT)~\cite{Hol86} and
effective meson Lagrangian methods~\cite{Mateu:2007tr}, which provide a
good description of low-energy meson phenomenology. However, one can also
address the question of how the form factors are connected to the
underlying quark structure. Due to the nonperturbative nature of the
quark-gluon interactions in the low-energy domain, to address this issue
one is forced to deal with models that treat quark interactions in some
effective way. In this sense, the widely studied Nambu--Jona-Lasinio (NJL)
model is the most popular schematic quark effective theory for
QCD~\cite{Vogl:1991qt,Klevansky:1992qe,Hatsuda:1994pi}. In this model
quarks interact through a local, chiral invariant four-fermion coupling.
Now, as a step towards a more adequate description of QCD, some extensions
of the NJL that include nonlocal interactions have been proposed in the
literature (see Ref.~\cite{Rip97} and references therein). In fact,
nonlocality arises as a natural feature of several well established
approaches to low energy quark dynamics, such as the instanton liquid
model~\cite{SS98}, the Schwinger-Dyson resummation
techniques~\cite{Roberts:1994dr}, and also lattice QCD
calculations~\cite{Parappilly:2005ei,Bowman2003,Furui:2006ks}. In the last
years nonlocal chiral quark models have been applied to study different
hadron observables with significant
success~\cite{Bowler:1994ir,Scarpettini:2003fj,GomezDumm:2006vz,Golli:1998rf,
Rezaeian:2004nf,Praszalowicz:2001wy,Noguera,Noguera:2005cc,Noguera:2008}.

In a previous work~\cite{Dumm:2010hh} we have addressed the analysis of the
vector and axial-vector form factors in $\pi^{+}\rightarrow
e^{+}\nu_{e}\gamma$ in the framework of nonlocal chiral quark models. We
have determined the values of $F_V(0)$, $F_A(0)$ and the slope of $F_V(q^2)$
at $q^2=0$ for different parameterization sets, comparing the results with
the experimental measurements and the corresponding values obtained in the
NJL model. In this paper we extend these previous results, analyzing the
vector form factor for values of $q^2$ up to 1 GeV$^2$, and studying the
slope of $F_A(q^2)$ at threshold. While the latter is interesting in view of
the comparison with future experimental results, the study of the behavior
of $F_V(q^2)$ presents relevant theoretical motivations. In fact, in the
isospin limit, and assuming the conserved-vector-current (CVC) hypothesis,
the vector form factor can be directly related to the form factor
$F^{\pi\gamma\gamma^{\ast}}(q^{2})$ associated with the vertex
$\pi\gamma\gamma^\ast$, where $q^2$ is the photon virtuality in Euclidean
space. Thus our results can be compared with the experimental information on
$F^{\pi\gamma\gamma^{\ast}}(q^{2})$, taken e.g.~from processes as $\pi^-
p\to \pi^0 n$ or $e^+ e^-\to e^+ e^-\pi^0$, and the compatibility with the
usually assumed single-pole behavior can be analyzed. The CVC hypothesis is
supported by the measured value of $F_{V}(0)$, which turns out to be fully
consistent with the corresponding value taken from
$F^{\pi\gamma\gamma^{\ast}}(q^{2})$ at $q^2 = 0$~\cite{pdg}. We notice that
the calculation of $F_V(q^2)$ in the framework of nonlocal models turns out
to be technically involved, due to the presence of poles and cuts arising
from the dressed quark propagators. Finally, to complete our analysis we
also study the compatibility of our nonlocal models with effective meson
theories such as $\chi$PT. In fact, it is seen that $F_A(0)$ and the pion
charge radius $\langle r^2\rangle_\pi$ are related to the so-called
$\chi$PT low energy constants (LECs) $\ell_5$ and $\ell_6$, which encode the
information of quark and gluon degrees of freedom. In $\chi$PT these LECs
are input parameters taken from phenomenology, thus the compatibility with
the predictions arising from effective quark models is a measure of the
ability of these models to account for the underlying strong interaction
physics.

The article is organized as follows: in Sect.~II we describe the models
and quote the analytical results for the vector form factor $F_V(q^2)$.
The numerical results for vector and axial-vector form factors are
presented and discussed in Sect.~III. In Sect.~IV we quote the values for
the LECs obtained within our models, while in Sect.~V we present our
conclusions. The full analytical expressions for the axial-vector form
factor are quoted in Appendix I.

\section{Formalism}

The amplitude for the process $\pi^{+}\rightarrow e^{+}\nu_{e}
(q)+\gamma(k)$ in Minkowski space can be written
as~\cite{Bryman:1982et}
\begin{align}
\mathcal{M}  &  = \frac{G_{F}}{\sqrt{2}}\ e\,\cos\theta_{C}\;\varepsilon^{\mu}
\left[  \sqrt{2}\;f_{\pi}\left\{  (q + k)^{\alpha}\,L_{\alpha\mu}\;-\;l^{\nu
}\left[  g_{\mu\nu}\;+\;\frac{q_{\mu}q_{\nu}}{(q\cdot k)}\right]  \right\}
\right. \nonumber\\
&  \left.  \qquad\qquad\qquad+ l^{\nu}\left\{  -\,i\,\epsilon_{\mu\nu
\alpha\beta}\;k^{\alpha}q^{\beta}\;\frac{F_{V}(q^{2})}{m_{\pi}}\;+\left[
q_{\mu}k_{\nu}\;-\;g_{\mu\nu}\,(q\cdot k)\right]  \frac{F_{A}(q^{2})}{m_{\pi}%
}\right\}  \ \right]  \ , \label{dos}%
\end{align}
with $(q+k)^{2}=m_{\pi}^{2}$, $k^{2}=0$. Here $G_{F}$ and $\theta_{C}$ stand
for the Fermi constant and the Cabibbo angle, respectively;
$\varepsilon_{\mu }$ is the photon polarization vector, $l_{\mu}$ is the
lepton current, $L_{\alpha\mu}$ is a lepton tensor, and $F_{V}(q^{2})$ and
$F_{A}(q^{2})$ denote the vector and axial-vector hadronic form factors
mentioned in the Introduction. In this work we are interested in the study
of these form factors in the context of nonlocal SU(2) chiral models that
include wave function renormalization. These models are defined by the
following Euclidean action \cite{Noguera,Noguera:2008}:
\begin{equation}
S_{E}=\int d^{4}x\ \left\{  \bar{\psi}(x)\left(  -i\rlap/\partial
+m_{c}\right)  \psi(x)-\frac{G_{S}}{2}\Big[j_{a}(x)j_{a}(x)+j_{P}%
(x)j_{P}(x)\Big]\right\}  \ . \label{action}%
\end{equation}
Here $m_{c}$ is the current quark mass, which is assumed to be equal for $u$
and $d$ quarks, while the nonlocal currents $j_{a}(x),j_{P}(x)$ are given by
\begin{align}
j_{a}(x)  &  =\int d^{4}z\ g(z)\ \bar{\psi}\left(  x+\frac{z}{2}\right)
\ \Gamma_{a}\ \psi\left(  x-\frac{z}{2}\right)  \ ,\nonumber\\
j_{P}(x)  &  =\int d^{4}z\ f(z)\ \bar{\psi}\left(  x+\frac{z}{2}\right)
\ \frac{i{\overleftrightarrow{\rlap/\partial}}}{2\ \varkappa_{p}}\ \psi\left(
x-\frac{z}{2}\right)  \ , \label{cuOGE}%
\end{align}
where $\Gamma_{a}=(\leavevmode\hbox{\small1\kern-3.8pt\normalsize1}, i
\gamma_{5}\vec{\tau})$ and $u(x^{\prime}) {\overleftrightarrow{\partial}}
v(x)=u(x^{\prime})\partial_{x}v(x) - \partial_{x^{\prime}}
u(x^{\prime})v(x)$. The functions $g(z)$ and $f(z)$ in Eq.~(\ref{cuOGE}) are
nonlocal covariant vertex form factors characterizing the corresponding
interactions. In what follows it is convenient to Fourier transform $g(z)$
and $f(z)$ into momentum space. Note that Lorentz invariance implies that
the corresponding Fourier transforms, $g_{p}$ and $f_{p}$, can only be
functions of $p^{2}$.

In order to deal with meson degrees of freedom, one can perform a standard
bosonization of the theory. This is done by considering the corresponding
partition function $\mathcal{Z}=\int\mathcal{D}\bar{\psi}\,\mathcal{D}%
\psi\,\exp[-S_{E}]$, and introducing auxiliary fields $\sigma_{1}%
(x),\sigma_{2}(x),\vec{\pi}(x)$, where $\sigma_{1,2}(x)$ and
$\vec{\pi}(x)$ are scalar and pseudoscalar mesons, respectively.
An effective bosonized action is obtained once the fermion fields
are integrated out. To treat that bosonic action we assume, as
customary, that $\sigma_{1,2}$ fields have nontrivial
translational invariant mean field values, while the mean field
values of pseudoscalar fields $\pi_{i}$ are zero. Thus we write
\begin{equation}
\sigma_{1}(x)=\bar{\sigma}_{1}+\delta\sigma_{1}(x)\ ,\qquad\sigma
_{2}(x)=\varkappa_{p}\ \bar{\sigma}_{2}+\delta\sigma_{2}(x)\ ,\qquad\vec{\pi
}(x)=\delta\vec{\pi}(x) \ .
\end{equation}
Replacing in the bosonized effective action and expanding in powers of meson
fluctuations we get
\begin{equation}
S_{E}^{\mathrm{bos}}\ =\ S_{E}^{\mathrm{MFA}}\ + \ S_{E}^{\mathrm{quad}}\ +
\ ...
\label{seucl}
\end{equation}
Here the mean field action per unit volume reads
\begin{equation}
\frac{S_{E}^{\mathrm{MFA}}}{V^{(4)}}=\frac{1}{2G_{S}}\left(  \bar{\sigma}%
_{1}^{2}+\varkappa_{p}^{2}\ \bar{\sigma}_{2}^{2}\right)  -4N_{c}\int
\frac{d^{4}p}{(2\pi)^{4}}\ \ln\left[  \frac{z_{p}}{-\rlap/p+m_{p}}\right]
^{-1}\ ,
\end{equation}
with
\begin{equation}
z_{p}=\left(  1-\bar{\sigma}_{2}\ f_{p}\right)  ^{-1}\ ,\qquad m_{p}
=z_{p}\left(  m_{c}+\bar{\sigma}_{1}\ g_{p}\right)  \ . \label{mz}%
\end{equation}
The minimization of $S_{E}^{\mathrm{MFA}}$ with respect to $\bar{\sigma}%
_{1,2}$ leads to the corresponding ``gap equations''. The quadratic terms in
Eq.~(\ref{seucl}) can be written as
\begin{equation}
S_{E}^{\mathrm{quad}}=\frac{1}{2}\int\frac{d^{4}p}{(2\pi)^{4}}\sum
_{M=\sigma,\sigma^{\prime},\pi}G_{M}(p^{2})\ \delta M(p)\ \delta M(-p)\ ,
\label{quad}%
\end{equation}
where $\sigma$ and $\sigma^{\prime}$ fields are scalar meson mass
eigenstates, defined in such a way that there is no $\sigma-\sigma^{\prime}$
mixing at the level of the quadratic action. The explicit expressions of
$G_{M}(p^{2})$, as well as those of the gap equations mentioned above, can
be found in Ref.~\cite{Noguera:2008}. Meson masses can be obtained by
solving the equations $G_{M}(-m_{M}^{2})=0$, while on-shell meson-quark
coupling constants $G_{Mq\bar{q}}$ are given by
\begin{equation}
\left(  G_{Mq\bar{q}}\right)  ^{-2}=\ \frac{dG_{M}(p^{2})}{dp^{2}%
}\bigg|_{p^{2}=-m_{M}^{2}}\ . \label{gpiqq}%
\end{equation}
As in Ref.~\cite{Noguera:2008}, we will consider here different functional
dependencies for the vertex form factors $g_{p}$ and $f_{p}$. First, we
consider a relatively simple case in which there is no wave function
renormalization of the quark propagator, i.e.~$f_{p}=0$, $z_{p}$ = 1, and we
take an often used exponential parameterization for $g_{p}$,
\begin{equation}
g_{p}=\mbox{exp}\left(  -p^{2}/\Lambda_{0}^{2}\right)  \ . \label{fg}%
\end{equation}
The model parameters $m_{c}$, $G_{S}$ and $\Lambda_{0}$ are determined by
fitting the pion mass and decay constant to their empirical values $m_{\pi
}=139$ MeV and $f_{\pi}=92.4$ MeV, and fixing the chiral condensate to the
phenomenologically acceptable value $\langle\bar{q}q\rangle^{1/3} = -240$
MeV. In what follows we refer to this choice of model parameters as set A.
Second, we consider a more general case that includes the $j_P(x)j_P(x)$
interaction, i.e.~we consider a nonzero wave function renormalization of
the quark propagator. We keep the exponential shape in Eq.~(\ref{fg}) for
the form factor $g_{p}$ and assume also an exponential form for $f_{p}$,
namely
\begin{equation}
f_{p}=\mbox{exp}\left(  -p^{2}/\Lambda_{1}^{2}\right)  \ . \label{ff}%
\end{equation}
Note that the range (in momentum space) of the nonlocality in each channel
is determined by the parameters $\Lambda_{0}$ and $\Lambda_{1}$,
respectively. The model includes now two additional free parameters,
namely $\Lambda_{1}$ and $\varkappa_p$. We determine here the five model
parameters so as to obtain the desired values of $m_{\pi}$, $f_{\pi}$ and
$\langle\bar{q}q\rangle^{1/3}$, and in addition we fix $m_c = 5.7$ MeV and
impose the condition $z_{p}(0)=0.7$, which is within the range of values
suggested by recent lattice
calculations~\cite{Parappilly:2005ei,Furui:2006ks}. We will refer to this
choice of model parameters and form factors as parameterization set B.
Finally, we consider a different functional form for the form factors,
given by
\begin{equation}
g_{p}=\frac{1+\alpha_{z}}{1+\alpha_{z}\ f_{z}(p)}\frac{\alpha_{m}%
\ f_{m}(p)-m\ \alpha_{z}f_{z}(p)}{\alpha_{m}-m\ \alpha_{z}}\ ,\qquad
f_{p}=\frac{1+\alpha_{z}}{1+\alpha_{z}\ f_{z}(p)}f_{z}(p)\ , \label{ff1}%
\end{equation}
where
\begin{equation}
f_{m}(p)=\left[  1+\left(  p^{2}/\Lambda_{0}^{2}\right)  ^{3/2}\right]
^{-1}\ ,\qquad f_{z}(p)=\left[  1+\left(  p^{2}/\Lambda_{1}^{2}\right)
\right]  ^{-5/2}\ . \label{parametrization_set2}%
\end{equation}
With this parameterization, the wave function renormalization of the quark
propagator and the effective quark mass have the simple expressions
$Z_{p}=1+\alpha_{z} f_{z}(p)$ and $m_{p}=m_{c}+\alpha_{m} f_{m}(p)$. As
shown in Ref.~\cite{Noguera:2008}, taking $m_{c}=2.37$ MeV, $\alpha
_{m}=309$ MeV, $\alpha_{z}=-0.3$, $\Lambda_{0}=850$ MeV and $\Lambda_{1}%
=1400$~MeV one can very well reproduce the momentum dependence of mass and
renormalization functions obtained in lattice calculations, as well as the
physical values of $m_{\pi}$ and $f_{\pi}$. We will refer to this choice
of model parameters as parameterization set~C. The parameter values for
all three parameter sets, as well as the corresponding predictions for
several meson properties, can be found in Ref.~\cite{Noguera:2008}.

In order to derive the form factors we are interested in, one should
\textquotedblleft gauge\textquotedblright\ the effective action $S_{E}$ by
introducing the electromagnetic field $A_{\mu}(x)$ and the charged weak fields
$W_{\mu}^{\pm}(x)$. For a local theory this \textquotedblleft
gauging\textquotedblright\ procedure is usually done by performing the
replacement
\begin{equation}
\partial_{\mu}\rightarrow\partial_{\mu}+i\ G_{\mu}(x)\ ,
\end{equation}
where
\begin{equation}
G_{\mu}(x)=\frac{e}{2}\ \left(  \frac{1}{3}+\tau^{3}\right)  A_{\mu}%
(x)+g_{W}\ \cos\theta_{C}\ \frac{(1-\gamma_{5})}{2}\ \frac{\tau^{+}W_{\mu}%
^{+}(x)+\tau^{-}W_{\mu}^{-}(x)}{\sqrt{2}}\ \ ,
\end{equation}
with $g_{W}^{2}/(8M_{W}^{2})=G_{F}\,/\sqrt{2}$ and $\tau^{\pm}=(\tau^{1}\pm
i\tau^{2})/2$. In the present case ---owing to the nonlocality of the involved
fields--- one has to perform additional replacements in the interaction terms,
namely
\begin{align}
\psi(x-z/2)\  &  \rightarrow\ W_{G}\left(  x,x-z/2\right)  \ \psi
(x-z/2)\ ,\nonumber\\
\psi^{\dagger}(x+z/2)\  &  \rightarrow\ \psi^{\dagger}(x+z/2)\ W_{G}\left(
x+z/2,x\right)  \ . \label{gauge}%
\end{align}
Here $x$ and $z$ are the variables appearing in the definitions of the
nonlocal currents [see Eq.(\ref{cuOGE})], and the function $W_{G}(x,y)$ is
defined by
\begin{equation}
W_{G}(x,y)\ =\ \mathrm{P}\;\exp\left[  i\ \int_{x}^{y}dr_{\mu}\ G_{\mu
}(r)\right]  \ , \label{intpath}%
\end{equation}
where $r$ runs over an arbitrary path connecting $x$ with $y$. Once the gauged
effective action is built, the explicit expressions for the vector and
axial-vector form factors can be obtained by expanding to leading order in the
product $\delta\pi^{+}\,A_{\mu}\,W_{\nu}^{+}$.

In the case of the vector form factor, one gets only a contribution
associated with the triangle diagram represented in Fig.~1a. This is given
by
\begin{equation}
\frac{F_{V}(q^{2})}{m_{\pi}^{+}}=-\frac{\sqrt{2}N_{c}}{3}\,G_{\pi q\bar{q}%
}\int\frac{d^{4}\ell}{(2\pi)^{4}}\ g_{\ell_{k-q}^{+}}\ \frac{(z_{\ell}%
+z_{\ell+k})z_{\ell-q}}{D_{\ell}\ D_{\ell+k}\ D_{\ell-q}}\ (t_{1}+t_{2})\ ,
\label{aaa}%
\end{equation}
where for convenience we have used the shorthand notation
$D_{\ell}=\ell^{2}+m_{\ell}^{2}$ and $\ell_{r}^{\pm}=\ell\pm r/2$. The terms
$t_1$ and $t_2$ in Eq.~(\ref{aaa}) read
\begin{align}
t_{1}  &  =\left(  \frac{1}{z_{\ell}}+\frac{1}{z_{\ell-q}}\right)  \left\{
\left[  \frac{\ell\cdot q}{k\cdot q}-q^{2}\frac{\ell\cdot k}{(k\cdot q)^{2}%
}\right]  m_{\ell+k}-\frac{\ell\cdot k}{k\cdot q}m_{\ell-q}\right. \nonumber\\
&  \qquad\qquad\qquad\qquad\qquad\qquad\left.  -\left[  1+\frac{\ell
\cdot(q-k)}{k\cdot q}-q^{2}\frac{\ell\cdot k}{(k\cdot q)^{2}}\right]
m_{\ell
}\right\}\ , \nonumber\\
& \nonumber\\
t_{2}  &  =\left[  \ell^{2}-2\frac{k\cdot\ell\ \ell\cdot q}{k\cdot q}%
+q^{2}\frac{(\ell\cdot k)^{2}}{(k\cdot q)^{2}}\right]  \left\{  \frac{1}%
{2}\left(  \frac{1}{z_{\ell}}+\frac{1}{z_{\ell-q}}\right)  \left(
\frac{m_{\ell}-m_{\ell-q}}{q\cdot\ell_{q}^{-}}+\frac{m_{\ell+k}-m_{\ell}%
}{k\cdot\ell}\right)  \right. \nonumber\\
&  \qquad\qquad\qquad\qquad\qquad\left.  +\frac{q^{2}}{q\cdot\ell_{q}^{-}%
}\left[  \bar{\sigma}_{1}\ \alpha_{g}^{+}(\ell_{q}^{-},q)+\frac{m_{\ell
-q}+m_{\ell}}{2}\bar{\sigma}_{2}\ \alpha_{f}^{+}(\ell_{q}^{-},q)\right]
\right\}\ , \label{t2}
\end{align}
with
\begin{equation}
\alpha_{h}^{\pm}(\ell,q)=\int_{0}^{1}\ d\lambda\ \frac{\lambda}{2}%
\ h_{\ell-\lambda\frac{q}{2}}^{\prime}\ \pm\ \int_{-1}^{0}\
d\lambda \ \frac{\lambda}{2}\ h_{\ell-\lambda\frac{q}{2}}^{\prime}
\ .
\label{alfa}%
\end{equation}
The integrals in $\alpha_{h}^{\pm}(\ell,q)$ have their origin in the path
dependence introduced through the ``gauged'' nonlocal effective action. The
result in Eqs.~(\ref{aaa}-\ref{alfa}) corresponds to the use of a straight
line between points $x$ and $y$ in Eq.~(\ref{intpath}).

For the case of the axial form factor the calculation is more involved since
it receives not only a contribution from the triangle diagram in Fig.~1a (as
occurs in the local NJL model) but also from other diagrams, which are
represented in Figs.~1b-1e. Since the resulting expressions are rather long,
we choose to relegate them to Appendix I where they are presented in detail.
We recall that we have been working in Euclidean space, therefore our
expressions for both $F_V(q^2)$ and $F_A(q^2)$ correspond to Euclidean
$q^2$. In order to obtain the form factors defined through the
$\pi^{+}\rightarrow e^{+}\nu_{e}\gamma$ amplitude in Minkowski space
[c.f.~Eq.~(\ref{dos})] one should change $q^2\to -q^2$.

\section{Numerical results for the vector and axial form factors}

In this section we present and discuss our numerical results for the
vector and axial-vector form factors. Let us start with the vector form
factor $F_{V}(q^{2})$. We will only restrict to values of $q^{2}$ which
are in the region where one can rely on the applicability of effective
chiral models, i.e.~$q^{2} \lesssim 1 \ \mbox{GeV}^{2}$. It is important
to stress that, even within that region, the numerical evaluation of
Eq.~(\ref{aaa}) presents some technical difficulties due to the presence
of poles in the corresponding integrand. Namely, once $q^{2}$ increases
beyond a certain threshold (about $0.1-0.2 \mbox{ GeV}^{2}$ for the
parameterizations considered here) the functions $D_{\ell+k}$ and
$D_{\ell-q}$ in Eq.(\ref{aaa}) can become zero for real values of
$\ell_{4}$. Thus, in order to properly evaluate the integrals, the
corresponding residues have to be conveniently added. Since for non-local
interactions the quark propagators might have several different poles,
this procedure has to be repeated each time a new pole threshold is
overpassed. In the case of set C the situation is even more complicated
due to the existence of a cut arising from the particular $q^{2}$
dependence of the interaction form factors, Eqs.~(\ref{ff1}) and
(\ref{parametrization_set2}). In any case, once these issues are taken
into account, a smooth $q^{2}$ dependence of the vector form factors is
obtained.

Before discussing in detail our predictions for the form factors
associated with the charged pion radiative weak decay, let us note that
one can also consider the related decay processes
$\pi^{0}\rightarrow\gamma\gamma$ and $\pi^{0}\rightarrow
e^{+}e^{-}\gamma.$ For these processes the amplitude involves the
$\pi^{0}\gamma\gamma$ vertex form factor $F^{\pi\gamma
\gamma^{\ast}}\left( q^{2}\right) $, where $q^{2}$ stands for the squared
invariant mass of the virtual photon, defined here in Euclidean space.
Experimental results can be well described by a single-pole parameterization,
\begin{equation}
F^{\pi\gamma\gamma^{\ast}}(q^{2})=\frac{F^{\pi\gamma\gamma^{\ast}}%
(0)}{1+a^{\prime}_{V}\ q^{2}/m_{\pi}^{2}}\ .
\label{gamma}
\end{equation}
Here it is important to mention that, in the chiral limit, the $U(1)$
axial anomaly leads to the constraint $F^{\pi\gamma \gamma^{\ast}}(0) =
1/(4\pi^{2}f_{\pi}) \simeq0.274 \ \mbox{GeV}^{-1}$, which is well
satisfied by the non-local models of the type considered
here~\cite{GomezDumm:2006vz}. Moreover, assuming the validity of the CVC
hypothesis together with isospin arguments it is possible to prove that
$F_{V}(0) = m_{\pi^{+}}\,F^{\pi \gamma\gamma^{\ast}}(0)/\sqrt{2}$. In the
case of the non-local models under consideration the validity of this
relation also extends to finite values of $q^{2}$, namely
\begin{equation}
F_{V}(q^{2}) = m_{\pi^{+}}\,F^{\pi\gamma\gamma^{\ast}}(q^{2})/\sqrt
{2}\ .
\label{relation}%
\end{equation}
The reason for this is that, in addition to respect CVC and isospin
symmetry, the models do not include explicit vector channels in the
quark-antiquark current-current interactions. As discussed below,
differences between the isoscalar and isovector vector channel interactions
could lead to violations of the relation in Eq.~(\ref{relation}) at finite
$q^{2}$.

Our results for $F_{V}(q^{2})$ are displayed in Fig.~2, where we plot the
curves corresponding to the form factors introduced in the previous
section. In addition, in Fig.~2 we show the results obtained from the
combination of Eq.~(\ref{relation}) with the parameterization in
Eq.~(\ref{gamma}). The parameters are taken from the experimental fit to
$\pi^{0} \to\gamma\gamma^{*}$ data, which yields
$F^{\pi\gamma\gamma^{\ast}}(0)=0.284(8)$ $\mbox{GeV}^{-1}$,
$a_{V}^{\prime}=0.032(4)$~\cite{pdg}. In general, it is seen that the
results for our parameter sets A, B and C turn out to be similar in the
whole range of $q^{2}$ values considered. The particular values at
$q^{2}=0$ are given in Table I. As already remarked in
Ref.~\cite{Dumm:2010hh}, one expect these values to be basically
coincident given the above mentioned constraint in the chiral limit.
Concerning the $q^{2}$ dependence, it is interesting to note that for all
three parameter sets the behavior can be very well reproduced by a
single-pole fit of the form
\begin{equation}
F_{V}(q^{2})=\frac{F_{V}(0)}{1+a_{V}\ q^{2}/m_{\pi}^{2}}\ .
\label{mono}%
\end{equation}
In view of the relation Eq.~(\ref{relation}), in our model $a_{V}$ can be
strictly identified with the parameter $a^{\prime}_{V}$ introduced in
Eq.~(\ref{gamma}).

On the other hand, it is possible to analyze the $q^2$ dependence of the
vector form factor considering the experimental data obtained from the
process $\pi^{+}\rightarrow e^{+}\nu_{e}\gamma$. For low values of $q^2$
one can carry out an expansion of the type
\begin{equation}
F_{V}(q^{2})=F_{V}(0)\,\left[  1-\lambda_{V}\,\frac{q^{2}}{m_{\pi^{+}}^{2}%
}+\lambda_{V}^{\prime}\,\frac{q^{4}}{m_{\pi^{+}}^{4}}\right] \ ,
\label{slope}
\end{equation}
where the minus sign arises from the definition of $\lambda_V = m_\pi^2
F_V'(0)/F_V(0)$ for $q^2$ in Minkowski space. In Table I the values of
$\lambda_{V},$ $\lambda _{V}^{\prime}$ and $a_{V}$ are given. We observe
that $a_{V}$ is slightly different from the value of $\lambda_{V}$. The
discrepancy can be taken as measure how much the actual $q^{2}$ dependence
of the vector form factors deviates from the single-pole behavior. This
slight deviation can be also seen from the closeness between $\lambda_V^2$
and $\lambda_V^\prime$, which should be equal in the single-pole limit. In
general, it is seen that for all three parameterizations the results for
$\lambda_V$ and $a_V$ are lower than experimental estimates (notice that
some enhancement is found when going from sets A, B to the
lattice-inspired set C). We remark that, given the fact that the
calculated values of $F_{V}(q^{2})$ are well adjusted by the single-pole
fit in Eq.~(\ref{mono}), the discrepancies between the model predictions
and the empirical fit observed in Fig.~2 can be traced back to the values
of $a_V$ (or $\lambda_V$), i.e.~the slopes of the curves at $q^2=0$. In
fact, these discrepancies are not unexpected, since our model does not
include vector-vector interactions. The magnitude of the corresponding
corrections can be estimated by considering e.g.\ the extended NJL model
studied in Ref.~\cite{Prades:1993ys}: indeed, it is seen that the vector
contribution to $\pi^{0}\rightarrow\gamma\gamma^{\ast}$ has the right
order of magnitude and sign in order to account for the discrepancies. It
is also interesting to point out that the contribution from the vector
channel could be different for $a_{V}$ and $a^{\prime}_{V}$, even when
isospin symmetry is preserved. This would be achieved if one has different
interactions in the vector-isoscalar channel and in the vector-isovector
channel. As a final comment concerning the vector form factor, let us
discuss the path dependence of our results. Contrary to what happens at
$q^{2}=0$, the values obtained at finite $q^{2}$ depend on the path
considered in Eq.~(\ref{intpath}). As usual, we have chosen a straight
line path for the calculations of the path dependent quantities
$\alpha_{h}^{\pm}(\ell,q)$, Eq.~(\ref{alfa}). To estimate the significance
of this path dependence we have performed single-pole fits to the
predictions obtained by neglecting the contribution of the the
corresponding terms in Eq.~(\ref{t2}). The resulting values for $a_{V}$
turn out to differ from the values listed in Table I by less than 0.5 \%.

We turn now to the axial form factor $F_{A}(q^{2})$. Contrary to the case of
the vector form factor, $F_{A}(q^{2})$ is not experimentally accessible for
(Euclidean) $q^{2}>0$. As stated, the value of $F_{A}(0)$ has been measured
from $\pi^{+}\rightarrow e^{+}\nu_{e}\gamma$ decay, and forthcoming
experiments offer the possibility of determining the slope $F_{A}^{\prime
}(0)= - F_{A}\left( 0\right) \,\lambda_{A}/m_{\pi^{+}}^{2}$. The model
predictions for these quantities are given in Table II. As it was discussed
in Ref.~\cite{Dumm:2010hh}, for $F_{A}(0)$ the triangle diagram in Fig.~1a
turns out to be the dominant one, giving at least 98\% of the total value
for all three parameter sets considered here. In the case of the slope the
triangle diagram is still the dominant one, but the contributions from
diagrams in Figs.~1b and 1e are also significant. One should note that
nonlocal models of the type considered here lead to values of $F_{A}(0)$
which are significantly different from those of $F_{V}(0)$. This is
remarkable, since other approaches like the NJL model~\cite{Courtoy:2007vy}
or the spectral quark model~\cite{Broniowski:2007fs} tend to give
$F_{V}(0) \simeq F_{A}(0)$. Given the triangle diagram dominance mentioned
above, the origin of this difference can be traced back to the different
dressing of the $\gamma_{\mu}$ and $\gamma_{\mu}\gamma_{5}$ terms in the
coupling of the $W$ to the quarks~\cite{Noguera:2008}. From Table II it is
seen that the predictions for $F_A(0)$ in nonlocal models are significantly
closer than the experimental value than the result obtained in the standard
NJL model. Regarding the slope parameter $\lambda_A$, we find that our
predictions are quite similar for the three parameter sets considered, and
turn out to be about 3/5 smaller than the value obtained in a resonance
effective model~\cite{Mateu:2007tr}.

\section{Predictions for LECs of Chiral Perturbation Theory}

In the previous section we have focused our attention on the
ability of our quark models to reproduce the main features of the
vector and axial-vector form factors. An alternative point of view
(see for example
Ref.~\citep{Schuren:1991sc}) is to consider quark models as
the generators of the pion $\chi$PT
Lagrangian~\citep{Gasser:1983yg}.
$\chi$PT describes the low energy physics of pions in a universal way, once the
order in the momentum and chiral breaking expansion (i.e.\ the order in
the chiral expansion) is specified. Different scenarios for quark models
will lead to $\chi$PT Lagrangians with different values of the
corresponding low energy constants (LECs).
In this section we analyze the connection between our
quark scenarios and the $\chi$PT Lagrangian up to the fourth order in the
chiral expansion.

The pionic Lagrangian that follows from performing a fourth order
gradient expansion of our model bosonized action in the presence
of external fields can be obtained using, for example, the method
described in Ref.~\cite{Schuren:1991sc}. On general grounds, it
should read
\begin{equation}
\mathcal{L}\ =\ \mathcal{L}_{2}+\mathcal{L}_{4}\ ,\label{ChiralExpansion}%
\end{equation}
where
\begin{eqnarray}
\mathcal{L}_{2} &  = &
\frac{f^2}{2} \ \nabla_{\mu}U^{T}\ \nabla^{\mu} U + \ f^2 \ \chi^T \ U
\ \ ,\label{L2Chiral}
\\
\mathcal{L}_{4} &  = &
\ell_{1}  \left( \nabla_{\mu} U^T \ \nabla^{\mu } U \right)^{2} +
\ell_{2}  \left( \nabla_{\mu} U^T \ \nabla_{\nu}U \right)
          \left( \nabla^{\mu} U^T \ \nabla^{\nu}U \right) +
\ell_{3}  \left( \chi^T U\right)^2 +
\nonumber\\
&  &
\ell_{4}  \left( \nabla_{\mu}\chi^T \ \nabla^{\mu}U \right)^{2}+
\ell_{5}  \  U^T F_{\mu\nu} F^{\mu\nu}U  +
\ell_{6}  \ \nabla^{\mu} U^T F_{\mu\nu}\nabla^{\nu} U +  ...
\ \ ,\label{L4Chiral}%
\end{eqnarray}
Here, $U=(s,p_i)$ is a four-component real
O(4) vector field of unit length, $U^T U=1$, and $\chi= 2 B (S,P_i)$ is
another O(4) vector proportional to the external scalar and pseudoescalar
fields. The constant $B$ is related to the vacuum expectation value of quarks
in the chiral limit and it can be expressed as $B= m^2/ 2 m_c$ where
$m$ is the pion mass at leading order in the chiral expansion.
This notation follows that of Ref.~\citep{Gasser:1983yg} where the definition
of the covariant derivative $\nabla_\mu$ and the field strength tensor
$F_{\mu\nu}$ can be also found. Note that among all possible terms in ${\cal L}_4$
only those relevant for our calculation have been explicitly given.

As a result of the procedure that leads to Eq.~(\ref{ChiralExpansion})
the explicit expressions for the LECs in terms of model parameters and
form factors $g_p$ and $f_p$ can be obtained and eventually numerically
calculated.
However, a simpler and more direct way to obtain their numerical
values is to extract them from the model predictions for the physical
quantities they are related to.
Of course, when proceeding in this way some care must be taken.
In fact, since the Lagrangian in Eq.~(\ref{ChiralExpansion})
is valid up to fourth order in the chiral expansion, to extract the
corresponding LECs from the physical quantities obtained in our
quark scenarios we should treat them to the same order of approximation.
Namely, we should calculate the relevant quantities in the chiral limit and
its vicinity.
As described in detailed in Ref.~\cite{Noguera:2008}, following this method
the LECs $\ell _{1},...,\ell_{4}$ can be obtained from the predicted $\pi-\pi$
scattering parameters. Thus, in what follows we will concentrate on the
remaining two parameters in Eq.~(\ref{L4Chiral}), namely $\ell_{5}$ and
$\ell_{6}$. These are related to the pion charge radius $\left\langle
r^{2}\right\rangle _{\pi}$ and $F_{A}\left(  0\right)  /m_{\pi}$
by~\cite{Gasser:1983yg}
\begin{align}
\ell_{5} &  =\frac{1}{4}\left(  \frac{f_{\pi}\,F_{A}\left(  0\right)  }%
{\sqrt{2}\,m_{\pi}}-\frac{1}{3}\left\langle r^{2}\right\rangle _{\pi}\,f_{\pi
}^{\,2}\right)\ ,  \nonumber\\
\ell_{6} &  =-\frac{1}{6}\left\langle r^{2}\right\rangle
_{\pi}f_{\pi}^{\,2}
\ .%
\end{align}
The required chiral limit values of $f_{\pi},$ $F_{A}\left(
0\right)  /m_{\pi}$ and $\left\langle r^{2}\right\rangle _{\pi}$
for our three parameterizations of nonlocal quark models are given
in Table III. The numerical results for $\ell_{5}$ and $\ell_{6}$
are given in Table IV, together with recent phenomenological
values obtained in $\chi$PT~\cite{GonzalezAlonso:2008rf} and the
values predicted within the Nambu-Jona Lasinio
model~\cite{Schuren:1991sc}. For completeness we have also
included the values of $\ell_{1},...,\ell_{4}$ from
Ref.~\cite{Noguera:2008}.

We observe that both the sign and the order of magnitude of the different
LECs are in reasonably good agreement with the results from $\chi$PT for
$\mu\sim 2m_\pi$. In fact, as already stressed in Ref.~\cite{Noguera:2008},
for Set C the constants $\ell_{2}$ and $\ell_{4}$ are particularly very well
reproduced, whereas the value of $\ell_{3}$ is acceptable given the existing
uncertainty. For $\ell_{1}$ we obtain a rather large range of values when
moving from one set of parameters to the other, but there is also a
significative variation of this LEC with the choice of $\mu$. Looking at the
two new LECs, $\ell_{5}$ and $\ell_{6}$, we find that the (absolute) values
obtained are relatively low, and the agreement with the phenomenological
values from $\chi$PT can be achieved only for somewhat larger values of the
renormalization point, $\mu\gtrsim m_\rho$. This can be understood taking
into account that our models do not include vector and axial-vector
currents. Therefore, even if the quark couplings lead to some interaction in
these channels, the models do not account for the effect of vector and
axial-vector meson resonances. This is consistent with the analysis of
Ref.~\cite{CERN-TH-5185/88}, which shows that $\ell_{5}$ and $\ell_{6}$ are
strongly dependent of the vector and axial-vector meson contributions. In
fact, previous works on the pion electromagnetic radius including resonances
indicate that vector meson contributions can give about 10$-$20\% of the
total result~\cite{nucl-th/9910033,hep-ph/0311359}.

\section{Conclusions}

We have studied the vector and axial-vector form factors in
$\pi^{+}\rightarrow e^{+}\nu_{e}\gamma$ in the framework of nonlocal chiral
quark models using three different sets of model parameters. As an extension
of a previous work~\cite{Dumm:2010hh} where we have determined the values of
$F_V(0)$, $F_A(0)$ and $F'_V(0)$, here we have analyzed the vector form
factor for values of $q^2$ up to 1 GeV$^2$ as well as the slope of
$F_A(q^2)$ at threshold. While the latter is interesting in view of the
comparison with future experimental results, the study of the behavior of
$F_V(q^2)$ presents relevant theoretical motivations, since it can be
related to the form factor $F^{\pi\gamma\gamma^{\ast}}(q^{2})$ associated
with the vertex $\pi\gamma\gamma^\ast$. We have found that the model
predictions for the $q^2$-dependence of the vector form factor can be well
described by a single-pole parameterization. This is indicated by the fact
that there is a slight difference between the values of the parameters $a_V$
and $\lambda_V$ for our three model parametrizations [the parameters $a_V$
and $\lambda_V$ are defined by Eqs.~(\ref{mono}) and (\ref{slope})]. On the
other hand, it is found that for all three parametrizations our results for
$a_V$ and $\lambda_V$ are below experimental estimates. These discrepancies
are not unexpected, since our model does not include vector-vector
interactions. In fact, a simple estimate based on an extended NJL
model~\cite{Prades:1993ys} indicates that the vector contribution to
$\pi^{0}\rightarrow\gamma\gamma^{\ast}$ has the right order of magnitude and
sign in order to account for the differences. This could be tested for the
present nonlocal models that include wave function renormalization through
the explicit inclusion of vector-vector interactions. Regarding the slope
parameter $\lambda_A$, we find that our predictions are quite similar for
the three parameter sets considered, and turn out to be about 3/5 smaller
than the value obtained in a resonance effective model~\cite{Mateu:2007tr}.
Finally, we also have extended the analysis of Ref.~\cite{Noguera:2008},
where the compatibility of our nonlocal models with effective meson theories
such as $\chi$PT is studied. Indeed, we have been able to extract the model
predictions for the $\chi$PT low energy constants (LECs) $\ell_5$ and
$\ell_6$, taking into account the relation between these LECs and the values
of $F_A(0)$ and the pion charge radius $\langle r^2\rangle_\pi$. Considering
the results for $\ell_5$ and $\ell_6$ together with the values of
$\ell_1,..,\ell_4$ previously obtained in Ref.~\cite{Noguera:2008}, we
observe that the predictions for both the sign and the order of magnitude of
the LECs within the present non-local models are in overall reasonably good
agreement with the results from $\chi$PT. For the particular case of
$\ell_{5}$ and $\ell_{6}$, we find that the (absolute) values obtained are
somewhat small, a fact that as in the case of $a_V$ and $\lambda_V$ might be
attributed to the lack of explicit vector and axial-vector interactions in
our models. In any case, it is seen that nonlocal model predictions for
$\ell_5$ are closer to the result of $\chi$PT than the value obtained in the
framework of the local NJL model.

\section*{Acknowledgements}

We would like to acknowledge useful discussions with J. Portol\'{e}s. This
work has been partially funded by the Spanish MCyT (and EU FEDER) under
contract FPA2010-21750-C02-01 and AIC10-D-000588, by Consolider Ingenio 2010
CPAN (CSD2007-00042), by Generalitat Valenciana: Prometeo/2009/129, by the
European Integrated Infrastructure Initiative HadronPhysics3 (Grant number
283286), by CONICET (Argentina) under grants \# PIP 00682 and PIP 02495, and
by ANPCyT (Argentina) under grant \# PICT07 03-00818.

\pagebreak

\section*{APPENDIX I: Explicit expressions for $F_{A}(q^{2})$}

In this appendix we present the expressions for the contributions of the
different diagrams in Fig.~1 to the axial form factor $F_{A}(q^{2})$. Let us
start by the triangle diagram in Fig.~1a, which yields
\begin{align}
\frac{F_{A}(q^{2})}{m_{\pi}}\Big|_{a}  &  = - N_{c}\,\sqrt{2}\ G_{\pi q\bar q}
\int\frac{d^{4}\ell}{\left(  2\pi\right)  ^{4}} \ g_{\ell^{+}_{k-q}}
\ \frac{z_{\ell+ k}\ z_{\ell- q} \ z_{\ell}} {D_{\ell+k}\ D_{\ell-q}
\ D_{\ell}} \left\{  A_{5} \left(  A \, c_{1} + B \, c_{2} + C\, c_{3}
\right)  \right. \nonumber\\
&  \left.  \qquad\qquad\qquad+ B_{5} \left(  A \, c_{4} + B \, c_{5} + C \,
c_{6} \right)  + C_{5} \left(  A \,c_{7} + B \, c_{8} + C \, c_{9} \right)
\right\}\ ,
\end{align}
where
\begin{align}
A  &  = \frac{1}{2} \left[  \frac{1}{z_{\ell+k}} + \frac{1}{z_{\ell}} \right]
\qquad; \qquad\!\!\!\!\!\! A_{5} =\frac{1}{2} \left[  \frac{1}{z_{\ell}} +
\frac{1}{z_{\ell-q}} \right] \nonumber\\
B  &  = \frac{1}{2 \ell\cdot k} \left[  \frac{1}{z_{\ell}} - \frac{1}%
{z_{\ell+k}}\right]  \ \ ; \qquad\!\!\!\!\!\!\!\! B_{5} = \frac{1}{2\ q
\cdot\ell_{q}^{-}} \left[  \frac{1}{z_{\ell-q}} - \frac{1}{z_{\ell}} \right]
+\frac{q^{2}\,\bar\sigma_{2}}{2\ q \cdot\ell_{q}^{-}} \alpha^{+}_{f}\left(
\ell^{-}_{q},q\right) \nonumber\\
C  &  = \frac{1}{\ell\cdot k} \left[  \frac{m_{\ell+k}}{z_{\ell+k}}%
-\frac{m_{\ell}}{z_{\ell}}\right]  \ \ ; \qquad\!\!\!\!\!\!\!\!\! C_{5}%
=\frac{1}{q \cdot\ell_{q}^{-}} \left[  2\frac{m_{\ell^{-}_{q}}}{z_{\ell
^{-}_{q}}}-\frac{m_{\ell}}{z_{\ell}}-\frac{m_{\ell-q}}{z_{\ell-q}}\right]
+\frac{q^{2}\,\bar\sigma_{1}}{q \cdot\ell_{q}^{-}} \alpha^{-}_{g}\left(
\ell^{-}_{q},q\right)\ .
\end{align}
Here, and in what follows, we use the already introduced shorthand notation
$\ell^{\pm}_{r} = \ell\pm r/2$, together with the functions $\alpha^{\pm}%
_{h}(\ell,q)$ defined in Eq.~(\ref{alfa}). The coefficients $c_{i}$ with
$i=1,...,9$ are given by
\begin{align}
c_{1}  &  = 2\, e_{1} \, \left(  m_{\ell+k} - m_{\ell-q} \right)  + 2\, e_{2}
\, \left(  m_{\ell}-m_{\ell+k}\right)  - 2\, e_{3} \, \left(  m_{\ell} +
m_{\ell-q}\right)  + 4 m_{\ell}\nonumber\\
c_{2}  &  = e_{1} \left[  -2 k\cdot\ell\left(  2 m_{\ell} + m_{\ell+k} -
m_{\ell-q}\right)  - 2 d_{1} + k\cdot q \left(  m_{\ell} + m_{\ell+k} \right)
\right] \nonumber\\
&  \qquad\qquad\qquad\qquad+ 2\,e_{3} \left[  d_{3} - \ell\cdot q \left(
m_{\ell}+m_{\ell+k}\right)  \right] \nonumber\\
c_{3}  &  = e_{1} \, \left[  k\cdot q - 2\ k\cdot\ell+d_{4} -d_{6}\right]  -
2\, e_{3}\left[  d_{5}- 2 \ell^{2} + \ell\cdot q \right] \nonumber\\
c_{4}  &  = e_{1} \left[  -2 d_{1} + 2 \ell\cdot q \left(  m_{\ell+k}%
+2m_{\ell}-m_{\ell-q}\right)  - q^{2} \left(  m_{\ell}+m_{\ell+k}\right)  +
k\cdot q \left(  m_{\ell}-m_{\ell-q}\right)  \right] \nonumber\\
&  \qquad+ 2\,e_{2} \left[  k\cdot\ell\left(  m_{\ell-q}-m_{\ell}\right)  +
d_{2} \right] \nonumber\\
c_{5}  &  = e_{1} \, \left[  \left(  4 k\cdot\ell+ 4\ \ell^{2}-4 \ell\cdot q -
k\cdot q \right)  \,d_{1} -2 \, k\cdot\ell\, d_{2}- 2\, \ell\cdot q \,
d_{3}\right. \nonumber\\
&  \left.  -2\, k\cdot\ell\, \ell\cdot q \left(  4 m_{\ell} + m_{\ell+k} -
m_{\ell-q} \right)  + q^{2} \, \left(  2\,\ell^{2} \left(  m_{\ell} +
m_{\ell+k}\right)  + k\cdot\ell\left(  3 m_{\ell} + m_{\ell+k} \right)
\right)  \right] \nonumber\\
c_{6}  &  = e_{1}\, \left[  k\cdot\ell\left[  2\,d_{5} - 2 \ell\cdot q +
q^{2}\right]  + \left(  2\ell^{2} - \ell\cdot q \right)  \, d_{6} - d_{4} \,
\left(  2\ell^{2}-3 \ell\cdot q + q^{2}\right)  - k\cdot q \,d_{5}\right]
\nonumber\\
c_{7}  &  = e_{1} \, \left[  d_{6} - d_{4}- 2\, \ell\cdot q - k\cdot q
\right]  + 2\, e_{2} \, \left[  d_{4} + 2 \ell^{2} + k\cdot\ell\right]
\nonumber\\
c_{8}  &  = e_{1} \, \left[  2 \, k\cdot\ell\left(  \ell\cdot q - d_{5}
\right)  - \left(  2\,\ell^{2} + k\cdot\ell\right)  \, d_{6} + \left(
2\,\ell^{2}-2 \, \ell\cdot q - k \cdot q + k \cdot\ell\right)  \, d_{4}
\right] \nonumber\\
c_{9}  &  = e_{1} \,\left[  d_{1} - k\cdot\ell\left(  m_{\ell-q}- m_{\ell}
\right)  - \ell\cdot q \left(  m_{\ell} + m_{\ell+k} \right)  - k\cdot q
\ m_{\ell}\right]\ ,
\end{align}
where
\begin{equation}%
\begin{array}
[c]{lcl}%
d_{1}= \ell^{2}\left(  m_{\ell} + m_{\ell+k}-m_{\ell-q}\right)  + m_{\ell-q}\,
m_{\ell} \,m_{\ell+k} & \ \ \ \ \ \ \ \ \ \  & d_{4}=m_{\ell} \, m_{\ell+k} -
\ell^{2}\\
d_{2}=\ell^{2}\left(  m_{\ell-q} + m_{\ell+k} - m_{\ell} \right)  + m_{\ell-q}
\, m_{\ell} \, m_{\ell+k} &  & d_{5}= m_{\ell} \, m_{\ell-q} + \ell^{2}\\
d_{3}=\ell^{2} \left(  m_{\ell} + m_{\ell+ k} + m_{\ell-q} \right)  -
m_{\ell-q} \, m_{\ell} \, m_{\ell+k} &  & d_{6}=m_{\ell-q}\,\left(  m_{\ell} +
m_{\ell+k} \right)
\end{array}
\end{equation}
and
\begin{eqnarray}
e_{1} = - \frac{ 6\;q^{2} \;(k\cdot\ell)^{2}}{(k\cdot q)^{3}} + \frac{8\;
k\cdot\ell\ \ell\cdot q}{(k\cdot q)^{2}} - \frac{2\;\ell^{2}}{k\cdot q} \ ;
\ e_{2} = -\frac{2 \; q^{2} \; k\cdot\ell}{(k\cdot q)^{2}}+ 2 \ \frac
{\ell\cdot q }{ k\cdot q} \ ; \ e_{3} = \frac{2 \; k\cdot\ell}{ k\cdot q }
\ .
\end{eqnarray}
The contributions of the other diagrams are the following:
\begin{eqnarray}
\frac{F_{A}(q^{2})}{m_{\pi}^{+}} \Big|_{b}  &  = & - \frac{\sqrt{2}}{2}\,G_{\pi
q\bar q}\,N_{c} \int\frac{d^{4}\ell}{\left(  2\pi\right)  ^{4}}\,\frac{e_{1}%
}{\ell\cdot k} \, \left(2 \, g_{\ell}- g_{\ell^{-}_{k}} - g_{\ell^{+}_{k}}
\right)  \frac{z_{\ell^{+}_{q}}}{D_{\ell^{+}_{q}}} \, \frac{z_{\ell^{-}_{q}}}%
{D_{\ell^{-}_{q}}} \Bigg\{ \frac{m_{\ell^{+}_{q}}}{z_{\ell^{-}_{q}}} -
\frac{m_{\ell^{-}_{q}}}{z_{\ell^{+}_{q}}}
\nonumber\\
& & + \left(  \frac{1}{z_{\ell^{+}_{q}}}- \frac{1}{z_{\ell^{-}_{q}}}\right)
\frac{m_{\ell^{+}_{q}}-m_{\ell^{-}_{q}}}{\ell\cdot q} \,\ell^{2} + \left(  2\,
\frac{m_{\ell}}{z_{\ell}}- \frac{m_{\ell^{+}_{q}}}{z_{\ell^{+}_{q}}}-
\frac{m_{\ell^{-}_{q}}}{z_{\ell^{-}_{q}}}\right)  \,\frac{ \ell^{+}_{q} \cdot
\ell^{-}_{q} + m_{\ell^{+}_{q}} \, m_{\ell^{-}_{q}}}{\ell\cdot q}
\nonumber\\
& & -\,\frac{q^{2}}{\ell\cdot q} \Bigg[ \bar\sigma_{1} \, \alpha^{-}_{g}\left(
\ell,q\right)  \left[  \ell^{+}_{q} \cdot\ell^{-}_{q} + m_{\ell^{+}_{q}} \,
m_{\ell^{-}_{q}} \right]\!  + \,\bar\sigma_{2} \, \alpha^{+}_{f} \left(  \ell,
q\right)  \, \ell\cdot\left[  \ell^{-}_{q} \, m_{\ell^{+}_{q}} - \ell^{+}_{q}
\, m_{\ell^{-}_{q}}\right]\! \Bigg] \Bigg\}
\nonumber\\
& &
\nonumber\\
\frac{F_{A} (q^{2})}{m_{\pi}^{+}}\Big|_{c}  &  = & \sqrt2 \ G_{\pi q\bar q}
\ N_{c} \int\frac{d^{4}\ell}{(2\pi)^{4}}\ \frac{ e_{1} }{\ell\cdot
k\ \ell\cdot q} \left[  \frac{z_{\ell_{k}^{-}} m_{\ell_{k}^{-}}}{D_{\ell
_{k}^{-}}} - \frac{z_{\ell_{k}^{+}} m_{\ell_{k}^{+}}}{D_{\ell_{k}^{+}}}
\right]  \left[  g_{\ell_{q}^{+}} + \frac{q^{2}}{2} \ \alpha^{+}%
_{g}(\ell,q) \right]
\nonumber\\
& &
\nonumber\\
\frac{F_{A}(q)}{m_{\pi^{+}}}\Big|_{d}  &  = & \sqrt{2} \ G_{\pi q\bar q} \ N_{c}
\int\frac{d^{4}\ell}{(2\pi)^{4}}\ \ \frac{e_{1}}{k \cdot\ell\ q \cdot\ell}
\left[  \frac{z_{\ell^{-}_{k}} \ m_{\ell^{-}_{k}} }{D_{\ell^{-}_{k}}}
-\frac{z_{\ell}\ m_{\ell}}{D_{\ell}}\right]
\nonumber\\
& & \qquad\qquad\qquad\qquad\times\left[  g_{\ell^{+}_{q}} - g_{\ell}%
-\frac{q^{2}}{2} \, \left(  \alpha^{+}_{g}(\ell,q) + \alpha^{-}_{g}(\ell,q)
\right)  \right]
\nonumber\\
& &
\nonumber\\
\frac{F_{A}(q^{2})}{m_{\pi}}\Big|_{e}  &  = & -\sqrt{2}\, G_{\pi q\bar q}\,
N_{c} \int\frac{d^{4}\ell}{\left(  2\pi\right)  ^{4}}\, g_{\ell}%
\frac{z_{\ell^{+}_{k+q}}}{D_{\ell^{+}_{k+q}}} \, \frac{z_{\ell^{-}_{k+q}}%
}{D_{\ell^{-}_{k+q}}} \Bigg\{ \bar\sigma_{1} \left[  \ell^{+}_{k+q}
\cdot\ell^{-}_{k+q} +m_{\ell^{+}_{k+q}} \, m_{\ell^{-}_{k+q}}\right]
\frac{\mathcal{\eta}_{g}\left(  \ell,k,q\right)  }{\left(  k\cdot q\right)
^{2}}
\nonumber\\[0.25cm]
& & -\bar\sigma_{2} \left[  m_{\ell^{-}_{k+q}} \ \ell\cdot\ell^{+}_{k+q} -
m_{\ell^{+}_{k+q}} \ \ell\cdot\ell^{-}_{k+q} \right]  \frac{\mathcal{\eta
}_{f}\left(  \ell,k,q\right)  }{\left(  k\cdot q\right)  ^{2}}%
\nonumber\\[0.25cm]
& & + \frac{1}{2\; k\cdot\ell^{+}_{q}} \,\left[  \frac{1}{z_{\ell^{+}_{q-k} }}
- \frac{1}{z_{\ell^{+}_{k+q}}} \right] \left[  m_{\ell^{+}_{k+q}} \left(
e_{1} +e_{2} - e_{3} - 1\right)  - m_{\ell^{-}_{k+q}} \left(  e_{1} +e_{2} +
e_{3} + 1\right)  \right]
\nonumber\\
& &
+ \frac{1}{2\; q\cdot\ell^{-}_{k}} \, \left[  \frac{1}{z_{\ell^{-}_{k+q}}%
}-\frac{1}{z_{\ell^{+}_{q-k}}} + q^{2}\,\bar\sigma_{2} \, \alpha^{+}_{f}
\left(  \ell^{-}_{k},q\right)  \right]  \left[  m_{\ell^{+}_{k+q}} \left(
e_{1} - e_{2} - e_{3} + 1\right)  \right.
\nonumber\\
& &
\left.  \qquad\qquad\qquad\qquad\qquad\qquad\qquad\qquad\qquad \ - \
m_{\ell^{-}_{k+q}} \left(  e_{1} + e_{2} - e_{3} - 1\right)  \right]  \Bigg\}
\end{eqnarray}
where%
\begin{align}
\eta_{h}(\ell,k,q)  &  = (k.q)^{2} \int_{0}^{1} d\lambda\int_{-1}^{0}%
d\lambda^{\prime}\ \left(  e_{1} + e_{2} \lambda+ e_{3} \lambda^{\prime}+
\lambda\lambda^{\prime}\right)  h^{\prime\prime}_{\ell^{+}_{\lambda q +
\lambda^{\prime}k}}
\end{align}
for $h=g$ or $f$.

\vspace*{1cm}
\begin{table}[ht]
\caption{Results for $F_{V}(q^{2})$ and its first and second derivative at
$q^{2}=0$. All results should be multiplied by $10^{-2}$. In column 5 we give
the empirical values of $m_{\pi^{+}}\,F^{\pi\gamma\gamma^{\ast}}(0)/\sqrt{2}$
and $a^{\prime}_{V}$. Note that in our model $a_{V}=a_{V}^{\prime}$. In column
6 we give the prediction of the local Nambu--Jona-Lasinio (NJL) model. }%
\label{tab1}
\begin{center}%
\begin{tabular}
[c]{cccccccc}\hline\hline
\hspace*{0.2cm}\ \hspace*{0.2cm} & \hspace*{0.2cm}Set A \hspace*{0.2cm} &
\hspace*{0.2cm}Set B \hspace*{0.2cm} & \hspace*{0.2cm}Set C \hspace*{0.2cm} &
\hspace*{0.2cm}Exp \cite{Bychkov:2008ws}\hspace*{0.2cm} & Exp ($\pi
^{0}\rightarrow\gamma\gamma^{\ast})$\cite{pdg} & \hspace*{0.2cm} NJL
\hspace*{0.2cm} & Ref \cite{Mateu:2007tr}\\\hline
$F_{V}(0)$ & 2.697 & 2.693 & 2.695 & 2.58(17) & 2.80(8) & 2.441 & 2.71\\
$a_{V}$ & 1.91 & 1.87 & 1.98 &  & 3.2(4) &  & \\
$\lambda_{V}$ & 1.651 & 1.726 & 2.011 & 10(6) &  & 3.244 & 4.1\\
$\lambda_{V}^{\prime}$ & 0.020 & 0.026 & 0.046 &  &  & $-$ & \\\hline\hline
\end{tabular}
\end{center}
\end{table}

\begin{table}[ht]
\caption{Results for $F_{A}(q^{2})$ and its first and second derivative at
$q^{2}=0$. All results should be multiplied by $10^{-2}$. In column 5 we give
the prediction of the local Nambu--Jona-Lasinio (NJL) model. }%
\label{tab2}
\begin{center}%
\begin{tabular}
[c]{ccccccc}\hline\hline
\hspace*{0.2cm}\ \hspace*{0.2cm} & \hspace*{0.2cm}Set A \hspace*{0.2cm} &
\hspace*{0.2cm}Set B \hspace*{0.2cm} & \hspace*{0.2cm}Set C \hspace*{0.2cm} &
\hspace*{0.2cm}Exp \cite{Bychkov:2008ws}\hspace*{0.2cm} & \hspace*{0.2cm} NJL
\hspace*{0.2cm} & From \cite{Mateu:2007tr}\\\hline
$F_{A}(0)$ & 1.319 & 1.614 & 1.825 & 1.17 $\pm$ 0.17 & 2.409 & exp. input\\
$\lambda_{A}$ & 1.22 & 1.17 & 1.26 &  & $-$ & 1.97\\
$\lambda_{A}^{\prime}$ & 0.012 & 0.013 & 0.034 &  & $-$ & \\\hline\hline
\end{tabular}
\end{center}
\end{table}

\begin{table}[ht]
\caption{Physical quantities used in the evaluation of $\ell_{5,6}$ obtained
in our different scenarios in the chiral limit.}
\begin{center}%
\begin{tabular}
[c]{ccccc}\hline\hline
&  & \hspace*{0.2cm} SA \hspace*{0.2cm} & \hspace*{0.2cm} SB \hspace*{0.2cm} &
\hspace*{0.2cm} SC\hspace*{0.2cm}\\\hline
$\left\langle r^{2}\right\rangle _{\pi}$ & fm$^{2}$ & 0.335 & 0.325 & 0.316\\
$F_{A}(0)/m_{\pi}$ & GeV$^{-1}$ & 0.097 & 0.118 & 0.133\\
$f_{\pi}$ & MeV & 91.2 & 91.4 & 91.8\\\hline\hline
\end{tabular}
\end{center}
\end{table}

\begin{table}[ht]
\caption{Values of $\ell_{i}$ obtained in our different scenarios. The $\chi
$PT values of $\ell_{i}^{r}$ as a function of $\mu$ are obtained from
Refs.~\cite{Colangelo:2001df,GonzalezAlonso:2008rf}. The last two columns
corresponds to the NJL predictions from Ref.~\cite{Schuren:1991sc} for two
different constituent quark mass:
$M=220,264\,\mathrm{MeV}$}%
.
\par
\begin{center}%
\begin{tabular}
[c]{ccccccccccc}\hline\hline
LEC & \multicolumn{3}{c}{Non Local QM} &  & \multicolumn{3}{c}{$\chi
$PT ($\ell_{i}^{r}(\mu)$)} &  & \multicolumn{2}{c}{NJL}\\
\cline{2-4}\cline{6-8}\cline{10-11}
$\times10^{3}$ & SA & SB & SC &  & $\mu=m_{\rho}$ & $\ \ \ \mu=2\ m_{\pi}$ &
$\ \ \mu=m_{\pi}$ &  & $M=220$ & $M=264$\\\hline
$\ell_{1}$ & -2.07 & -1.39 & 0.26 &  & $-4.0\pm0.6$ & $-1.9\pm0.6$ &
$-0.4\pm0.6$ &  & -0.63 & -2.3\\\hline
$\ell_{2}$ & 6.51 & 6.46 & 6.41 &  & $1.9\pm0.2$ & $6.2\pm0.2$ & $9.1\pm0.2$ &
& 6.3 & 6.2\\\hline
$\ell_{3}$ & -1.1 & -2.3 & -4.1 &  & $1.5\pm4.0$ & $-1.8\pm4.0$ & $-4.0\pm4.0$
&  & -8.5 & -3.5\\\hline
$\ell_{4}$ & 15.0 & 17.2 & 20.3 &  & $6.2\pm1.3$ & $19.1\pm1.3$ & $27.9\pm1.3$
&  & 22.7 & 12.2\\\hline
$\ell_{5}$ & -4.39 & -3.90 & -3.54 &  & $~-5.22\pm0.06$ & $~-6.29\pm0.06$ &
$~-7.02\pm0.06$ &  & -2.88 & -2.60\\
$\ell_{6}$ & -11.9 & -11.6 & -11.4 &  & $-13.1\pm0.4$ & $-15.3\pm0.4$ &
$-16.8\pm0.4$ &  & -11.5 & -10.4\\
$\left(  2\,\ell_{5}-\ell_{6}\right)$ & 3.15 & 3.82 & 4.33 &  & $2.7\pm0.4$
& $2.7\pm0.4$ & $2.7\pm0.4$ &  & 5.75 & 5.21\\\hline\hline
\end{tabular}
\end{center}
\end{table}

\vfill

\pagebreak

\begin{figure}[ht]
\includegraphics[width=0.8 \textwidth]{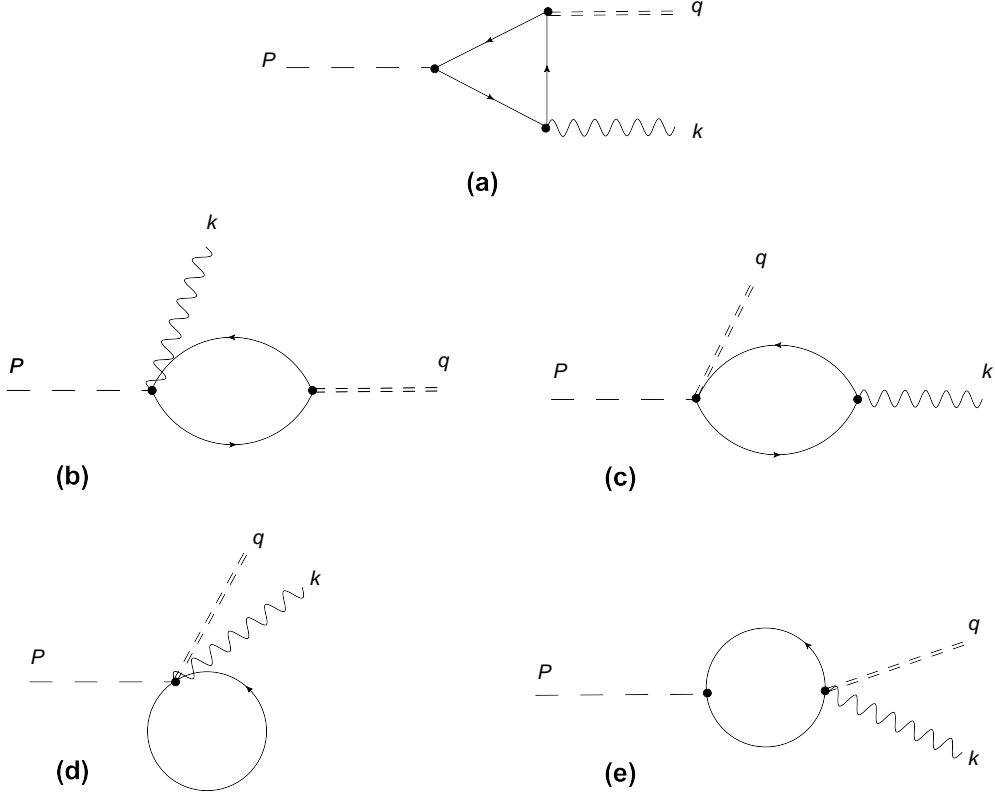}\caption{Diagrammatic
representation of the possible contributions to $\pi^{+}\rightarrow e^{+}%
\nu_{e}\gamma$ decay. Double-dashed lines, wavy lines and single-dashed lines
represent the $e\,\nu_{e}$ pair, the outgoing photon and the decaying pion,
respectively. While for the vector form factor only the contribution from the
triangle diagram (a) is nonvanishing, in the case of the axial-vector form
factor all five diagrams contribute.}%
\label{diaI}%
\end{figure}

\begin{figure}[ht]
\includegraphics[width=0.8 \textwidth]{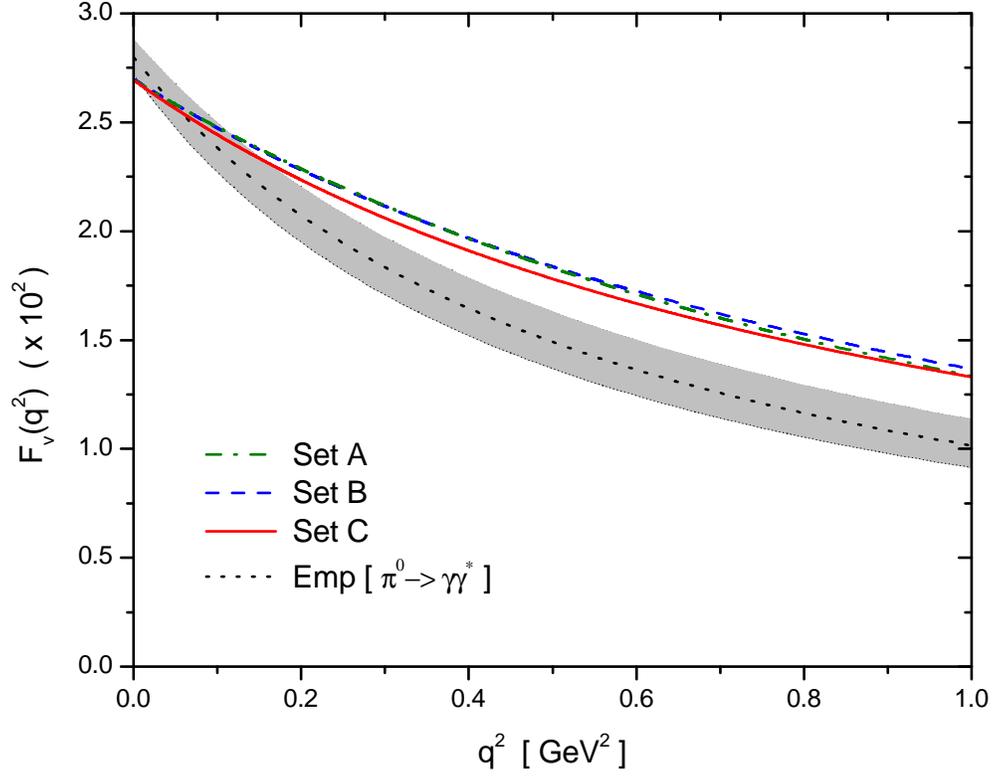}
\caption{$F_{V}(q^2)$ for sets $A$, $B$ and $C$. The shadowed region
corresponds to empirical data on $F^{\pi\gamma\gamma^\ast}$, using
Eqs.~(\ref{gamma}) and (\ref{relation})}.
\label{fig2}%
\end{figure}

\end{document}